\documentclass[nofootinbib,twocolumn,showkeys,showpacs,preprintnumbers,
               prd,amsmath,amssymb,aps,floats,floatfix]{revtex4}

\usepackage{mathrsfs}
\usepackage{epsfig}
\usepackage{graphicx}
\usepackage{dcolumn}
\usepackage{bm}
\usepackage[usenames]{color}
\usepackage{tabularx}
\usepackage[caption=false]{subfig}
\usepackage{color}

\usepackage{multirow}
\usepackage{wrapfig}
\usepackage{booktabs}
\usepackage[T1]{fontenc} 
\usepackage{mathrsfs}
\usepackage{color}
\usepackage{slashed}
\usepackage{enumerate}
\usepackage{helvet}
\usepackage{bigstrut}
\usepackage{array}

\usepackage[normalem]{ulem}

\definecolor{darkgreen}{cmyk}{1,0,1,0.4}

\definecolor{darkred}{cmyk}{0,1,0.5,0.4}

\def\ie{{\em i.e. }}

\newcommand{\ba}{\begin{array}}
\newcommand{\ea}{\end{array}}
\newcommand{\bd}{\begin{displaymath}}
\newcommand{\ed}{\end{displaymath}}
\def\be{\begin{equation}}
\def\ee{\end{equation}}
\def\bsube{\begin{subequation}}
\def\esube{\end{subequation}}
\def\bea{\begin{eqnarray}}
\def\eea{\end{eqnarray}}
\def\bal{\begin{align}}
\def\ealign{\end{align}}
\def\eal{\end{align}}
\def\ben{\begin{enumerate}}
\def\een{\end{enumerate}}
\def\nn{\nonumber}

\def\beq{\begin{equation}\ba{rcl}}
\def\eeq{\ea\end{equation}}
\def\d{\partial}

\def\gev{\; {\rm GeV} }
\def\tev{\; {\rm TeV} }

\def\bsub{\begin{subequations}}
\def\esub{\end{subequations}}

\def\b{\beta}
\def\e{\epsilon}

\def\L{\Lambda}

\newcommand{\gonev}{\ensuremath{g^\text{\em\tiny V}_{_1} } }
\newcommand{\kv}{\ensuremath{\kappa_{_V} } }
\newcommand{\lv}{\ensuremath{\lambda_{_V}} }

\newcommand{\lgam}{\ensuremath{\lambda_{_\gamma}} }
\newcommand{\lz}{\ensuremath{\lambda_{_Z}} }

\newcommand{\dkg}{\ensuremath{\Delta \kappa_{_\gamma}} }
\newcommand{\dkz}{\ensuremath{\Delta\kappa_{_Z}} }
\newcommand{\dczeroz}{\ensuremath{\Delta c^\text{\em\tiny ZZ}_{_0}} }
\newcommand{\dczerow}{\ensuremath{\Delta c^\text{\em\tiny WW}_{_0}} }

\newcommand{\dczerozg}{\ensuremath{\Delta c^{\text{\em\tiny Z}\gamma}_{_0}} }
\newcommand{\dczerog}{\ensuremath{\Delta c^{^{\gamma\gamma}}_{_0}} }
\newcommand{\dcoz}{\ensuremath{\Delta c^\text{\em\tiny ZZ}_{_1}} }
\newcommand{\dcow}{\ensuremath{\Delta c^\text{\em\tiny WW}_{_1}} }

\newcommand{\dcozg}{\ensuremath{\Delta c^{\text{\em\tiny Z}\gamma}_{_1}} }
\newcommand{\dcog}{\ensuremath{\Delta c^{^{\gamma\gamma}}_{_1}} }
\newcommand{\dgoz}{\ensuremath{\Delta g^\text{\em\tiny Z}_{_1}} }

\newcommand{\daoz}{\ensuremath{\Delta a^\text{\em\tiny ZZH}_{_1}}}
\newcommand{\daow}{\ensuremath{\Delta a^\text{\em\tiny WWH}_{_1}} }
\newcommand{\atwoz}{\ensuremath{a^\text{\em\tiny ZZH}_{_2} } }
\newcommand{\atwow}{\ensuremath{a^\text{\em\tiny WWH}_{_2} } }
\newcommand{\daog}{\ensuremath{\Delta a^\text{\em\tiny $\gamma\gamma$H}_{_1}} }
\newcommand{\daozg}{\ensuremath{\Delta a^\text{\em\tiny Z$\gamma$ H}_{_1}} }
\newcommand{\atwozg}{\ensuremath{a^\text{\em\tiny Z$\gamma$H}_{_2} } }
\newcommand{\atwog}{\ensuremath{a^\text{\em\tiny $\gamma \gamma$H}_{_2} } }

\newcommand{\daozh}{\ensuremath{\Delta a^\text{\em\tiny ZZHH}_{_1}}}
\newcommand{\daowh}{\ensuremath{\Delta a^\text{\em\tiny WWHH}_{_1}} }
\newcommand{\atwozh}{\ensuremath{a^\text{\em\tiny ZZHH}_{_2} } }
\newcommand{\atwowh}{\ensuremath{a^\text{\em\tiny WWHH}_{_2} } }
\newcommand{\daogh}{\ensuremath{\Delta a^\text{\em\tiny $\gamma\gamma$HH}_{_1}} }
\newcommand{\daozgh}{\ensuremath{\Delta a^\text{\em\tiny Z$\gamma$HH}_{_1}} }
\newcommand{\atwozgh}{\ensuremath{a^\text{\em\tiny Z$\gamma$HH}_{_2} } }
\newcommand{\atwogh}{\ensuremath{a^\text{\em\tiny $\gamma \gamma$HH}_{_2} } }
\newcommand{\dbohc}{\ensuremath{\Delta b_{\text{\tiny 1}}^{\text{\tiny $H^3$}} }}
\newcommand{\dbohq}{\ensuremath{\Delta b_{\text{\tiny 1}}^{\text{\tiny $H^4$}} }}
\newcommand{\btwohc}{\ensuremath{b_{\text{\tiny 2}}^{\text{\tiny $H^3$}} }}
\newcommand{\btwohq}{\ensuremath{b_{\text{\tiny 2}}^{\text{\tiny $H^4$}} }}

\newcommand{\rts}{\ensuremath{\sqrt{s}} }

\newcommand{\tthw}{\ensuremath{\tan{\theta_\text{\em\tiny \!W}}} }
\newcommand{\sthw}{\ensuremath{\sin{\theta_\text{\em\tiny \!W}}} }
\newcommand{\cthw}{\ensuremath{\cos{\theta_\text{\em\tiny \!W}}} }
\newcommand{\tthws}{\ensuremath{\tan^2{\!\theta_\text{\em\tiny \!W}}} }
\newcommand{\sthws}{\ensuremath{\sin^2{\!\theta_\text{\em\tiny \!W}}} }

\newcommand{\cthws}{\ensuremath{\cos^2{\!\theta_\text{\em\tiny \!W}} }}
\newcommand{\secthws}{\ensuremath{\sec^2{\!\theta_\text{\em\tiny \!W}} }}

\newcommand{\ct}{\ensuremath{\cos{\theta}} }
\newcommand{\st}{\ensuremath{\sin{\theta}} }
\newcommand{\sth}{\ensuremath{{\text s}_{_\theta}} }
\newcommand{\cth}{\ensuremath{{\text c}_{_\theta}} }
\newcommand{\mz}{\ensuremath{m_{_Z} } }
\newcommand{\mzsq}{\ensuremath{m_{_Z}^{2} } }

\newcommand{\mw}{\ensuremath{m_{_W} } }
\newcommand{\mwsq}{\ensuremath{m_{_W}^{2} } }
\newcommand{\mws}{\ensuremath{m_{_W}^{2} } }
\newcommand{\mh}{\ensuremath{m_{_H} } }
\newcommand{\mhsq}{\ensuremath{m_{_H}^{2} } }

\newcommand{\ordershalf}{\ensuremath{{\cal O}\!\left(\sqrt{s}/m\right)} }
\newcommand{\orders}{\ensuremath{{\cal O}\!\left(s/m^2\right)} }
\newcommand{\ordersthalf}{\ensuremath{{\cal O}\!\left(s^{3/2}/m^3\right)} }
\newcommand{\orderss}{\ensuremath{{\cal O}\!\left(s^2/m^4\right)} }

\newcommand{\fphith}{f_{\phi 3}^{}}

\makeatletter
\DeclareMathSizes{\@xpt}{\@xpt}{5}{5}
\makeatother

\begin{document}
\title{Investigating Perturbative Unitarity in Presence of Anomalous Couplings}
\author{Mamta Dahiya$^a$}\email{mamta.phy26@gmail.com}
\author{Sukanta Dutta$^a$}\email{sukanta.dutta@gmail.com}
\author{Rashidul Islam$^{b}$}\email{islam.rashid@gmail.com}
\affiliation{$^a$SGTB Khalsa College, University of
Delhi. Delhi-110007. India.}
\affiliation{$^b$Department of Physics, University of Calcutta,\\ 92, Acharya Prafulla Chandra Road, Kolkata 700009, India.}
\begin{abstract}
 We perform a model independent analysis  of the helicity  amplitudes at high energy for all the $2\to2$ scattering processes involving gauge and Higgs bosons in the presence of anomalous $WWV$, $WWVV$, $VVH$, $VVHH$ ($V\equiv Z,\gamma$  and $W^\pm$),  $HHHH$ and $HHH$    interactions. We obtain the perturbative unitarity constraints on anomalous couplings by demanding the vanishing of terms  proportional to $s^2$ and $s^{3/2}$ in the helicity amplitudes. Using these constraints, we also compute the upper bound on all the anomalous couplings from terms linear  in $s$.
\par Further, assuming all anomalous couplings to have arisen only from  dimension six operators, we show that the perturbative unitarity violation can be evaded up to $\sim$ 9 TeV
corresponding to the best fit values of $f_{WW}/\Lambda^2$ and $f_{BB}/\Lambda^2$
from the combined analysis of Tevatron and LHC data.
\end{abstract}
\maketitle
\section{Introduction}
\label{sec:intro}
With the discovery of a new massive ($\sim 125\gev$) scalar particle by both ATLAS
and CMS \cite{HiggsDiscovery} and with most of the observations and
consistency checks indicating that the new particle has a large overlap with  the Higgs boson of the Standard Model (SM)
\cite{HiggsProp}, it is now possible to directly explore the fabric of the
electroweak symmetry breaking mechanism. In order to find out how the $SU(2)_L
\times U(1)_Y$ symmetry is broken in nature, one needs to measure precisely the
strength of the self-interactions of the Higgs boson and its interactions with the gauge
bosons as well as the fermions.  Although there are enough reasons to expect the presence of  new
physics beyond SM, a good agreement of the SM predictions with the experiments so far
imply that any new model must reduce to the SM at low energies. Thus even if the SM is only a low energy effective theory valid upto some energy scale $\Lambda$, the
observation of any departure from the SM predicted values of  the interaction strengths in  the bosonic sector 
can give hints of  new physics (NP).  The specific form of the NP
which will supersede the SM is not yet known. However, a model independent approach can
be adopted  either to observe the  signatures of the NP spectrum, if any, though too heavy  to be produced in the
present colliders, or realize their effects in the precision
measurements of the novel interactions among the known SM particles {\it via} loop corrections.
There are two conceptual model independent  prescriptions. One approach to an effective field theory is to extend the theory by adding higher dimensional operators constructed out of the SM fields. The other approach is by writing  Lagrangian containing all possible Lorentz structures that can contribute to a given process but with fewest number of derivatives~\cite{eff-lag,Degrande:2012wf}.

With the present data, we can safely assume that the observed new scalar state
belongs indeed to a light electroweak doublet scalar and that the $SU(2)_L
\times U(1)_Y$ symmetry is linearly realized in the effective theory. The effect of any NP at energy below the cut-off scale can be parametrized as effective interactions in a theory whose particle
content is the same as in the SM \cite{eff-lag}. In the effective field theory (EFT) framework
\cite{Degrande:2012wf}, operators constructed out of the SM fields and of dimension higher than four are added to
the SM Lagrangian\footnote{In the SM Lagrangian, all operators are restricted
to be of mass dimension four or less.}. These higher dimensional
operators are suppressed by appropriate powers of the cut-off scale $\Lambda$.
\begin{align}
 {\cal L}_{\rm eff}
 =
 {\cal L}_{\rm SM}
 +
 \sum_{n > 4} \sum_i \frac{f_i^{(n)}}{\L^{(n-4)}} {\cal O}_i^{(n)},
 \label{fulllagrangianop6}
\end{align}
where ${\cal L}_{\rm SM}$ denotes the renormalizable SM Lagrangian and
${\cal O}_i^{(n)}$s are the gauge invariant operators of mass dimension $n$.
The index $i$ runs over all operators (consistent with
the symmetries of the SM) of the given mass dimension, and the coefficients $f_i^{(n)}$ are
dimensionless parameters, which are determined once the full theory is known. The scale $\Lambda$ can be regarded as the scale of new physics and is large compared
with the experimentally-accessible energies. Thus the dominant extended operators will be those of the lowest dimensionality and hence we will be concerned with the dimension six operators in this article. Recently attempt has been made   to constrain  the coefficients of dimension six operators with the Higgs boson data from LHC in references \cite{Ellis:2014jta,Falkowski:2015fla}.

In the anomalous coupling approach, the most general effective Lagrangian is
written assuming  Lorentz invariance, Bose symmetry and $SU(3)_C$ and electromagnetic gauge invariance. This might contain additional Lorentz structures that are not present in the SM Lagrangian. The deviation from 
the SM value of the coefficient corresponding  to a given  Lorentz structure induces an  
anomalous coupling, which has no inherent  scale dependence. Usually, an effective 
Lagrangian contains all possible Lorentz structures, each constructed with the fewest 
number of derivatives. One can construct an infinite number of additional terms by adding 
derivatives~\cite{Hagiwara:1986vm} but to be conservative, one  retains the terms containing least number of derivatives.  However, with constant anomalous couplings, unitarity is broken at some scale and to circumvent this problem sometimes arbitrary momentum dependent form factors are introduced in the vertex. This is the momentum space analogue of the infinite number of terms
in the Lagrangian approach that can be constructed by including more derivatives. 

Thus, the effective Lagrangian in  both approaches contain infinite number of terms and if the entire series are considered, both approaches would be equivalent. 

We attempt to address the problem of preserving perturbative unitarity of all $2 \to 2$ scattering processes in the gauge and the Higgs boson sector in the presence of anomalous couplings. We shall also relate the analysis with the dimension six operators.

With just the gauge sector Feynman diagrams, the $VV$ scattering amplitudes within the SM
grow with energy and eventually violate unitarity. If the symmetry breaking is due to a light
Higgs boson, the Higgs Mechanism removes this famous bad high energy
behaviour and restores unitarity. The nuances of non-Abelian gauge structure of SM ensures the cancellation
of order $s^2$ terms ($\rts$ being the centre of mass energy) among the gauge mediated diagrams while order $s$ terms cancel among the gauge and Higgs boson mediated diagrams and hence the perturbative unitarity is preserved.
 Thus any appreciable deviation in the $s$ dependence of the scattering amplitudes from
 that within the SM provides a rather sensitive test of
the anomalous couplings in high energy $VV$, $VH$ or $HH$ scattering experiments. 

\par Recently the authors of reference ~\cite{Corbett:2014ora} performed a similar
analysis as ourselves after the preliminary version of our article had appeared
in the arXiv~\cite{Dahiya:2013uba}. Taking a cue from the study of
reference ~\cite{Csaki:2003dt} they  dropped the most dominant helicity amplitudes
which are either proportional to $s^2$ or $s^{3/2}$  assuming
that these terms will be automatically cancelled by demanding $SU(2)_L\times U(1)_Y$
gauge invariant sum rules and consequently arrived at  the  perturbative unitarity conditions with  terms linearly proportional to $s$ only. However, it is important to note that the sum rules derived  in reference  \cite{Csaki:2003dt} for Higgsless models with infinite KK modes and also in recently reviewed article on the bulk Higgs models  in reference \cite{Csaki:2015hcd} are neither valid for the SM  nor for the SM with finite  number of light Higgs models. As a consequence, we observe that, a priori, these sum rules do not hold true for   dimension six operators involving   light Higgs bosons irrespective of   whether they are $SU(2)_L\times U(1)_Y$ gauge invariant or not. Thus, the helicity amplitudes that grow with the  centre of mass energy as $s^n$ ( $n\ge 0$ ) do not get cancelled automatically.

\par In this
article, we investigate the high energy behaviour of the 
scattering amplitudes for the following sixteen distinct scattering processes:  $W^+W^-\to W^+W^-$,  $W^+W^-\to Z(\gamma)Z(\gamma)$, $W^+W^-\to Z\gamma$, $ZZ \to Z(\gamma)Z(\gamma)$, $Z(\gamma)Z(\gamma)\to Z \gamma$, $\gamma \gamma \to \gamma \gamma $, $W^
+W^- \to Z(\gamma )H$, $W^+W^-$ $\rightarrow HH$, $Z(\gamma)Z (\gamma)\to HH$,  $Z\gamma\to HH$, $HH \to HH$ in the presence of
 the anomalous trilinear gauge,  quartic gauge, the Higgs--gauge boson and
 Higgs self couplings\footnote{Note that the helicity amplitudes of other
 scattering processes  such as $W^\pm W^\pm \to W^\pm W^\pm $, $W^\pm
 Z (\gamma) \to W^\pm  Z(\gamma)$ and $W^\pm  Z(\gamma) \to W^\pm \gamma
 (Z)$  are related to the ones mentioned here by crossing symmetry. Thus
 the high energy behaviour of these follow the same suit as for the
 processes analysed in the article.}.

In particular, we ask if it is possible to preserve perturbative unitarity even in the
presence of anomalous couplings in the gauge and Higgs sector and determine the values of these anomalous
couplings allowed by unitarity constraints for all the scattering processes considered.
Vector boson scattering has drawn a lot of attention earlier and many
works exist in the literature that discuss the gauge boson scattering and
unitarity problems \cite{ww-scatt}. The anomalous gauge and gauge-Higgs couplings and
their limits in various models as well as in model-independent approach have
 also been studied in various papers~\cite{Hagiwara:1996kf,Dutta:2008bh,Hagiwara:1993ck,Corbett:2012dm,Kumar:2014zra,Liu-Sheng:2014gxa,Dolan:2013rja,Banerjee:2013apa,Cheung:2010af}.

The rest of the paper is organised as follows: We provide the framework
of our calculations in Section~\ref{sec:formalism}. In  Section~\ref{sec:partial}  we discuss  the high energy behaviour of the
partial wave amplitudes for various scattering processes mentioned above and obtain the unitarity constraints on the linear combination of  anomalous couplings.
 In Section~\ref{sec:d6op}, we relate all the anomalous couplings to the coefficients of
dimension six operators. A summary of our results is given in Section~\ref{sec:conc}. The notation of our calculations are given in the Appendix ~\ref{app:notation}.
\section{Formalism--Anomalous Couplings}
\label{sec:formalism}
Within the SM, the interactions among the bosons of the electroweak
theory are determined entirely  by the gauge symmetry. Any deviations from the
SM couplings are, therefore, evidences of new physics. As mentioned in
previous Section, these deviations from the SM predictions may be parametrized
in a model independent way in terms of effective Lagrangian. The terms of the
effective Lagrangian relevant for the processes considered by us may be written
as
\begin{align}
 {\cal L}_{\rm eff}
 =&
 {\cal L}_{\rm eff}^{WWV}
 + {\cal L}_{\rm eff}^{WWVV^\prime}
 + {\cal L}_{\rm eff}^{V_1V_2H}
 + {\cal L}_{\rm eff}^{V_1V_2HH}
 \nn\\
 &
 + {\cal L}_{\rm eff}^{H^3}
 + {\cal L}_{\rm eff}^{H^4},
 \label{lag:pheno}
\end{align}
where ${\cal L}_\text{eff}^{WWV}$ and ${\cal L}_\text{eff}^{WW VV^\prime}$ 
gives rise to  triple gauge couplings (TGC) and quartic gauge couplings (QGC)  involving two $W$ bosons respectively. ${\cal L}^{V_1V_2H}_{\rm eff}$  and
 ${\cal L}^{V_1V_2H H}_{\rm eff}$ lead to anomalous vertices
involving the Higgs boson and electroweak gauge bosons while ${\cal L}^{H^3}_{\rm eff}$ and ${\cal L}^{H^4}_{\rm eff}$ respectively generate cubic and quartic Higgs 
self couplings. Note that we are not considering effective Lagrangian involving
only (three or four) neutral gauge bosons as these are absent at tree level in SM 
and also (in anticipation) because these are not generated by 
dimension six operators as we shall see in Section~\ref{sec:d6op}.

\par
Restricting our study to CP-even vertices only, the triple gauge vertices involving two $W$ bosons can be parametrized as~\cite{Hagiwara:1986vm}
\begin{align}
 {\cal L}_{\rm eff}^{WWV}
 =&
 {\rm i} g_{WWV}
 \Big[ \gonev \left( W^+_{\mu\nu} W^{-\mu} V^\nu - W^+_\mu V_\nu W^{-\mu\nu} \right)
 \nn\\
 &
 + \kv W^+_\mu W^-_\nu V^{\mu\nu}
 + \frac{\lv}{\mwsq} W^-_{\mu\nu} W^{+\nu\rho} V_{\ \rho}^\mu \Big],
 \label{eq:lagwwv}
\end{align}
with $W_{\mu\nu} = \partial_\mu W_\nu - \partial_\nu W_\mu$ and $V_{\mu\nu} =
\partial_\mu V_\nu - \partial_\nu V_\mu$. Also, $g_{_{WW\gamma}} = -\,e= -\,g\sthw$ and
$g_{_{WWZ}}=-\,e \cot{\theta_W} = -\,g\cthw$, $\theta_W$
being the weak mixing angle. In the SM, at tree level, $\gonev =
\kv =1$ and $\lv = 0$. Writing each TGC as sum of the SM part and the anomalous part, these vertices involve six C and P conserving anomalous
couplings but demanding the electromagnetic gauge invariance requires that
$g_{1}^{\gamma} = 1$ leaving five anomalous TGCs, namely $\dgoz \equiv g_{_1}^\text{\em\tiny Z} - g_{_{1,{\text  SM}}}^\text{\em\tiny Z} $, $ \dkg \equiv \kappa_{_\gamma} -\kappa_{_{\gamma , {\text SM}}}$, $\dkz \equiv \kappa_{_Z} -\kappa_{_{Z, {\text SM}}}$, \lgam and \lz. 
The constraints on the TGC from LEP  \cite{Schael:2013ita} and LHC \cite{Chatrchyan:2013yaa}   are obtained by assuming the relations 
\begin{gather}
  \lz = \lgam = \lambda,
  \label{eq:cond-gaugeinv2} \\
  \dkz = \dgoz - \dkg \tthws,
  \label{eq:cond-gaugeinv3}
\end{gather}
We shall also assume these constraints to be valid in our analysis in next Section.

\par The effective interactions of four electroweak gauge bosons
(QGC), may be
parametrized in terms  of two Lorentz invariant structures, given by the Lagrangian \cite{Eboli:2006wa}
\begin{eqnarray}
 {\cal L}_\text{eff}^{WW V V^\prime}
 = & \,
 c^{V V^\prime}_0 {\cal O}^{V V^\prime}_0
 - c^{V V^\prime}_1 {\cal O}^{V V^\prime}_1,
\label{qgcstruc-lag}
\end{eqnarray} 
\vskip -0.2cm
where
\vskip -0.4cm
\begin{align}\label{qgcstruc-op}
\begin{split}
 {\cal O}^{WW}_0 =& \frac{1}{2} g^{\alpha\gamma} g^{\beta\delta}
                   [W^+_\alpha W^+_\beta W^-_\gamma W^-_\delta],
 \\
  {\cal O}^{WW}_1 =& \frac{1}{2} g^{\alpha\beta} g^{\gamma\delta}
                   [W^+_\alpha W^+_\beta W^-_\gamma W^-_\delta],
 \\
 {\cal O}^{VV}_0 =& g^{\alpha\gamma} g^{\beta\delta}
                   [W^+_\alpha W^-_\beta V_\gamma V_\delta];\,\, V = Z/\gamma,
 \\
 {\cal O}^{VV}_1 = &g^{\alpha\beta} g^{\gamma\delta}
                   [W^+_\alpha W^-_\beta V_\gamma V_\delta];\,\, V = Z/\gamma,
 \\
 {\cal O}^{Z\gamma}_0 =& (g^{\alpha\gamma} g^{\beta\delta} + g^{\alpha\delta} g^{\beta\gamma})
                   [W^+_\alpha W^-_\beta Z_\gamma A_\delta],
 \\
  {\cal O}^{Z\gamma}_1 =& 2 g^{\alpha\beta} g^{\gamma\delta}
                   [W^+_\alpha W^-_\beta Z_\gamma A_\delta].
\end{split}
\end{align}
The SM  tree level quartic couplings associated with these Lorentz structures are given as
\bea
- c^\text{\em\tiny WW}_{_{0,{\text SM}}} &=& -c^\text{\em\tiny WW}_{_{1,{\text SM}}} = c^\text{\em\tiny ZZ}_{_{0,{\text SM}}}/\cthws = c^\text{\em\tiny ZZ}_{_{1,{\text SM}}}/\cthws \nn\\
&=& c^{^{\gamma\gamma}}_{_{0,{\text SM}}}/\sthws = c^{^{\gamma\gamma}}_{_{1,{\text SM}}}/\sthws \nn\\
&=&  c^{\text{\em\tiny Z$\;\!\gamma$}}_{_{0,{\text SM}}}/(\cthw\sthw) =  c^{\text{\em\tiny Z$\;\!\gamma$}}_{_{1,{\text SM}}}/(\cthw\sthw) \nn\\
&=& g^2.\label{eq:qgc-sm}
\eea
Accordingly we can define   eight anomalous  quartic gauge  couplings (AQGC) as $\Delta c^{VV^\prime}_{_i} \equiv c^{V V^\prime}_{_i}-c^{V V^\prime}_{_{i,{\text SM}}}$ with  $i=0,1$, corresponding  to all Lorentz structures listed in~\eqref{qgcstruc-op}. It is to be noted that, to restrict our parameter space, we are not considering 
the Lorentz structures involving derivatives of the gauge fields. Further, we shall not consider AQGC $\dczerog$ and $\dcog$ which means that we assume the couplings of two photon fields with two the $W$ fields to be same as in the SM. The reason will be clear in Section~\ref{sec:d6op} when we see that these are not generated by the dimension six operators considered by us.

\par Similarly, demanding only Lorentz invariance, the most general form of CP-even coupling
between  a pair of gauge bosons  ($V_1$ and $V_2$) and the Higgs boson is given
by \cite{Choudhury:2006xe}
\begin{eqnarray}
&&\Bigl(\Gamma_{\text{\tiny eff}}^\text{\tiny $V_1 V_2 H$}\Bigr)_{\mu\nu}
\,\,=\nn\\
&& g \mw \Bigl[ a_{\text{\tiny 1}}^\text{\tiny $V_1 V_2 H$} g_{\mu\nu}
 + \frac{ a_{\text{\tiny 2}}^\text{\tiny $V_1 V_2 H$}}{\mzsq} \bigl( p_{_{2\mu}} p_{_{1\nu}} - g_{\mu\nu} \, p_{_1} \cdot p_{_2} \bigr) \Bigr] \, .
\label{eq:vertex-hvv} 
\end{eqnarray}
where $V_1V_2H$  corresponds to  $\gamma\gamma H$, $ Z \gamma H$, $ZZH$ and $W^+ W^-H$ vertices.

Identically  the most general Lorentz invariant CP-even couplings
between  a pair of gauge bosons  ($V_1$ and $V_2$) and a pair of Higgs bosons may be parametrised as
 \begin{eqnarray}
&&\Bigl(\Gamma_{\text{\tiny eff}}^\text{\tiny $V_1 V_2 H H$}\Bigr)_{\mu\nu}
\,\,=\nn\\
&& \frac{g^2}{2}  \Bigl[ a_{\text{\tiny 1}}^\text{\tiny $V_1 V_2 H H$} g_{\mu\nu}
 + \frac{ a_{\text{\tiny 2}}^\text{\tiny $V_1 V_2 H H$}}{\mzsq} \bigl( p_{_{2\mu}} p_{_{1\nu}} - g_{\mu\nu} \, p_{_1} \cdot p_{_2} \bigr) \Bigr]\, .
\label{eq:vertex-hhvv} 
\end{eqnarray}
In \eqref{eq:vertex-hvv} and \eqref{eq:vertex-hhvv}, $p_{_1}$, $p_{_2}$ are the incoming momenta of the two gauge bosons. At tree level
in SM, we have
\bea a_{\text{\tiny 1, SM}}^\text{\tiny\em ZZH} &=& a_{\text{\tiny 1, SM}}^\text{\tiny\em ZZHH} =\sec^2{\theta_W}/2,\nn\\ 
a_{\text{\tiny 1, SM}}^\text{\tiny\em WWH} &=& a_{\text{\tiny 1, SM}}^\text{\tiny\em WWHH} =1 ,\nn\\ 
 a_{\text{\tiny 1, SM}}^\text{\tiny$ Z\gamma H$} &=& a_{\text{\tiny 1, SM}}^\text{\tiny$ \gamma\gamma H$} =a_{\text{\tiny 1, SM}}^\text{\tiny$ Z\gamma HH$} = a_{\text{\tiny 1, SM}}^\text{\tiny$ \gamma\gamma H H$} =0 \, ,\nn\\
{\text{and}}\quad a_{\text{\tiny 2, SM}}^\text{\tiny $V_1 V_2 H$}&=& a_{\text{\tiny 2, SM}}^\text{\tiny $V_1 V_2 H H$}=0 \,.
\label{eq:vvh-vvhh-sm}
\eea

 Thus, writing $a_{\text{\tiny 1}}^\text{\tiny $V_1 V_2 H(H)$} = a_{\text{\tiny 1, SM}}^\text{\tiny $V_1 V_2 H(H)$}(1+ \Delta  a_{\text{\tiny 1}}^\text{\tiny $V_1 V_2 H(H)$})$,
 we have eight anomalous $VVH$ (four $\Delta  a_{\text{\tiny 1}}^\text{\tiny $V_1 V_2 H$}$   and four  $a_{\text{\tiny 2}}^\text{\tiny $V_1 V_2 H$}$) couplings and   similarly eight $V_1V_2HH$ couplings.
However, the present and low energy data indicates 
that the effective Lagrangian should better  preserve the SM gauge symmetries, which then  requires that the tree level SM prediction for $VVH$ and $VVHH$ may  not be  modified. Hence we take
\begin{gather}
 \daoz = \daow = \daog = \daozg = 0,
 \label{eq:cond-gaugeinv1} \\
 \daozh = \daowh = \daogh = \daozgh = 0.
 \label{eq:cond-gaugeinv1a} 
\end{gather}
reducing the total number of anomalous $VVH$ and $VVHH$ couplings to eight.
\par The Higgs boson self coupling measurements and their deviations from the SM expectations provide the hint of alternative scenarios of symmetry breaking with light Higgs boson. These interactions can be  parametrised  as follows~\cite{Barger:2003rs} 
\begin{eqnarray}
{\cal L}_{\text{\tiny eff}}^\text{\tiny $H^3$}
&=&  - \frac{m^2_H}{2 v} \left[ \left( 1 + \dbohc \right) H^3 - 3 \frac{\btwohc}{m^2_H} H (\d_\mu H) (\d^\mu H) \right],\nn\\
 \label{eq:lag-selfh1} \\
{\cal L}_{\text{\tiny eff}}^\text{\tiny $H^4$}
 &=&- \frac{m^2_H}{8 v^2} \left[ \left( 1 + \Delta b_{\text{\tiny 1}}^{\text{\tiny $H^4$}}  \right) H^4 - 6 \frac{\btwohq}{m^2_H} H^2 (\d_\mu H) (\d^\mu H) \right].
\nn\\
\label{eq:lag-selfh2} 
\end{eqnarray}
In this parametrisation, there are four anomalous Higgs boson self couplings, namely, \dbohc, \dbohq, \btwohc and \btwohq all of which are zero in the SM at the tree level.

\par Considering all anomalous couplings to be constant and perturbative unitarity be preserved upto a
given energy scale, we attempt to compute
the upper bound on all the anomalous couplings discussed in this Section. Since we assume the Higgs boson to be SM-like, all couplings at tree level are assumed
to be close to their SM values. 
\section{Partial Wave Analysis}
\label{sec:partial}

\begin{figure}[!ht]
  \centering
\includegraphics[width=0.47\textwidth]{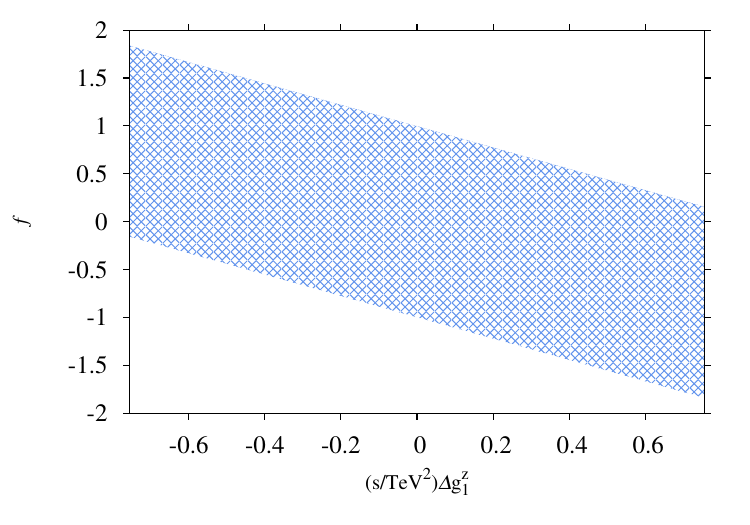}
\caption{\small\em The shaded enclosed region on the plane of $f$ and \dgoz corresponds to the constraints from ${\cal A}^0_{0,0,0,0}(WW\to ZZ)$, ${\cal A}^0_{0,0,\pm,\pm}(ZZ\to Z\gamma)$ and  ${\cal A}^0_{0,0,\pm,\pm}(WW\to Z\gamma)$.} 
\label{fig:bound2}
\end{figure}
\begin{figure}[!ht]
  \includegraphics[width=0.47\textwidth]{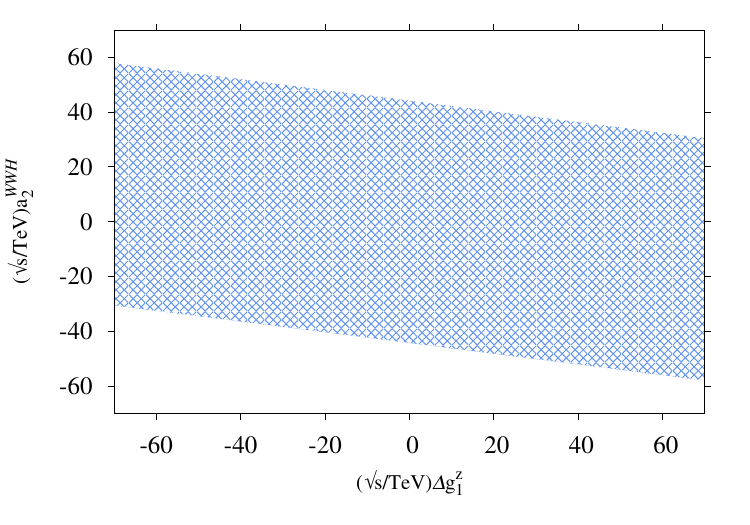}
\caption{\small\em The shaded region is an infinite band on the $(\rts\atwow-\rts\dgoz)-$ plane constrained by inequation given in \eqref{eq:bound3}.}
\label{fig:bound3}
\end{figure}
Partial wave analysis of scattering processes is one of the often used methods
to constrain unknown parameters in a theory \cite{ww-scatt}. For a given $2 \to 2$ scattering  process $a(p_a,\,\lambda_a) + b (p_b,\,\lambda_b) \to 
c(p_c,\,\lambda_c) + d (p_d,\,\lambda_d) $, the invariant transition amplitude
${\cal M}_{fi}$ can be decomposed in terms of partial wave amplitudes 
${\cal A}^J_{\lambda_a\, \lambda_b\,\lambda_c\,\lambda_d}(s) $ as~\cite{spin-formalism} 
\bea 
{\cal M}_{fi}(s,\, \Omega) =
16 \pi \sum_J ( 2J+1) {\cal A}^J_{\lambda_a\, \lambda_b\,\lambda_c\,\lambda_d}(s) 
D^{J*}_{\lambda \lambda^\prime}(\phi,\theta,0) \nn\\
\label{eq:def-partialamp}
\eea
where $\lambda=\lambda_a-\lambda_b$, $\lambda^\prime=\lambda_c-\lambda_d$, 
 and  $\Omega \equiv (\theta,\phi) $ is the solid angle. $D^J_{\lambda\lambda^\prime}$ is the standard rotation matrix and we have chosen
 (in the c.m. frame), $\vec p_a =-\vec p_b = \vec p_i = |\vec p| \hat z$
 while  $\vec p_c =-\vec p_d = \vec p_f$ to be along direction $(\theta,\phi) $.
The partial wave amplitudes may be obtained from transition amplitude ${\cal M}_{fi}$
by inverting equation~\eqref{eq:def-partialamp} and using orthogonality relation of 
rotation matrices as
\bea
 {\cal A}^J_{\lambda_a\, \lambda_b\,\lambda_c\,\lambda_d}(s) = 
 \frac{1}{64\pi^2} \int d \Omega\,\, D^{J}_{\lambda \lambda^\prime}(\phi,\theta,0){\cal M}_{fi}(s,\, \Omega). \nn\\
\label{eq:partialamp}
\eea 

The unitarity of S-matrix which is equivalent to $T^\dagger - T = i T^\dagger T$, 
with $S=I + iT$, requires that even the most dominant partial amplitude ${\cal A}^J$ should satisfy
\bea
\left\vert {\cal R} e \left( {\cal A}^J(s) \right)\right\vert \le 1/2.
\label{eq:unitarity-cond0}
\eea
For a given $J$, the high energy behaviour (\ie behaviour at energies much higher than $m$, the mass of the heaviest particle involved in the scattering process) of the amplitude ${\cal A}^J(s)$ may be studied by expanding the amplitudes in powers of $s/m^2$ as
\begin{eqnarray}
{\cal A}^J(s) =\sum_{n=-\infty}^\infty c_n^J\,\,\left[\frac{s}{m^2}\right]^{n/2}\, .
\end{eqnarray} 
Terms generated for $n<0$ approach zero at high energies (i.e. for $\rts \gg m$). Terms for $n >0$ grow with energy and hence the unitarity condition will require either the coefficient $c_n^J$ to vanish or satisfy
\bea
\displaystyle\sum_{n>0} c_n^J\,\,\left[s/m^2\right]^{n/2}\le \frac{1}{2} \quad \text{for} \quad \rts \gg m. 
\label{eq:unitarity-cond}
\eea
\begin{widetext}
\begin{center}
\begin{table}[ht]\footnotesize
\begin{tabular}{cllll||cl}
\hline
\hline
 Anomalous & \qquad Unitarity Bound& in unit of & \hskip 0.5 cm Associated &\qquad & Anomalous  & Unitarity \\
couplings &  & $1\tev^2/s$&\hskip 0.5 cm Partial  Amplitude(s) &\qquad &  Couplings  &Bound in unit \\
 &  & & &\qquad &  related to $\dgoz$ &  of $1\tev^2/s$ \\
\hline
&&&& &&\\
$|\dgoz|$ &$\qquad\leq \,\frac{4\sthws}{\alpha} \Bigl(\frac{\mws}{s} \Bigr)$
& $\simeq \, 0.756 $ & \hskip 0.5 cm 
${\cal A}^0_{0000}({WW\to ZZ})$ &\qquad \eqref{eq:wwzz0000}&& \\ 
&&&&&$\quad \,\left\vert\dkz \right\vert\,$ &$\le$ 0.756\\
&&&&&&\\
&&&&&&\\
$|\dcozg|$ &$\qquad\leq \,\frac{2\tthw}{\alpha} \Bigl(\frac{\mws}{s} \Bigr)$ & $\simeq \, 0.898 $  & \hskip 0.5 cm
${\cal A}^0_{00\pm\pm}({WW\to Z\gamma})$ & \qquad \eqref{eq:wwzg00mm} 
& $ \left\vert\dczeroz \right\vert$&$\le$ 1.51\\ 
&&&&&&\\
& & & \hskip 0.5 cm ${\cal A}^0_{\pm\pm00}({ZZ\to ZZ})$ &\qquad\eqref{eq:zzzz-00mm}&&\\
 $|\atwoz|$ &$\qquad \leq \,\frac{8\tthws}{\alpha} \Bigl(\frac{ \mws}{s} \Bigr)$ & $\simeq\, 1.97$ & \hskip 0.5 cm
${\cal A}^0_{00\pm\pm}({ZZ\to ZZ})$ &\qquad\eqref{eq:zzzz-00mm}
& $\left\vert\dcoz \right\vert$&$\le$ 1.51\\
&& &\hskip 0.5 cm ${\cal A}^0_{00\pm\pm}({WW\to ZZ})$&\qquad \eqref{eq:wwzz-00mm}&&\\
&&&&&&\\
 $|\atwog|$ & $\qquad\leq \,\frac{8\tthws}{\alpha} \Bigl(\frac{\mws}{s} \Bigr)$ & $\simeq  \, 1.97$ &\hskip 0.5 cm 
${\cal A}^0_{00\pm\pm}({ZZ\to \gamma\gamma})$ &\qquad\eqref{eq:zzgg-00mm}
&$\left\vert\dczerozg \right\vert$&$\le$ 0.756 \\
   $|\atwozg|$ &$\qquad\leq \,\frac{8\tthws}{\alpha} \Bigl(\frac{\mws}{s} \Bigr)$
& $\simeq \, 1.97 $ &
$ \hskip 0.5 cm {\cal A}^0_{00\pm\pm}({ZZ \to Z\gamma})$  &\qquad\eqref{eq:zzzg-m00m}& & \\ 
&&&&&&\\
& & & \hskip 0.5 cm $ {\cal A}^0_{\pm\pm 00}({WW\to WW})$ &\qquad\eqref{eq:wwwwmm00}
&$\left\vert\dczerow \right\vert $ &$\le$ 1.16\\
 $|\atwow|$ & $\qquad\leq \,\frac{8\tthws}{\alpha} \Bigl(\frac{\mws}{s} \Bigr)$
& $\simeq \, 1.97 $ 
&$ \hskip 0.5 cm {\cal A}^0_{00\pm\pm}({WW\to WW})$ & \qquad\eqref{eq:wwwwmm00} &&\\ 
& & &\hskip 0.5 cm $ {\cal A}^0_{\pm\pm 00}({WW\to ZZ})$ &\qquad\eqref{eq:wwzzmm00}
&$\left\vert\dcow\right\vert $&$\le$ 1.51\\ 
&&&&&&\\
 $|\atwowh|$ & $\qquad\leq \,\frac{8\tthws}{\alpha} \Bigl(\frac{\mws}{s} \Bigr)$
& $\simeq \, 1.97 $ & \hskip 0.5 cm
${\cal A}^0_{\pm,\pm}({WW\to HH})$ & \qquad\eqref{eq:wwhhmm} &&\\ 
&&&&&&\\
 $|\atwozh|$ & $\qquad\leq \,\frac{8\tthws}{\alpha} \Bigl(\frac{\mws}{s} \Bigr)$
& $\simeq  \, 1.97$ & \hskip 0.5 cm
 ${\cal A}^0_{\pm,\pm}({ZZ\to HH})$ & \qquad\eqref{eq:zzhhmm} && \\ 
&&&&&&\\
 $|\atwozgh|$ &  $\qquad\leq \,\frac{8\tthws}{\alpha} \Bigl(\frac{\mws}{s} \Bigr)$
 & $\simeq\, 1.97  $ & \hskip 0.5 cm
 ${\cal A}^0_{\pm,\pm}({Z\gamma\to HH})$ &\qquad \eqref{eq:zghhmm} &&  \\ 
&&&&&&\\
 $|\atwogh|$ &  $\qquad\leq \,\frac{8\tthws}{\alpha} \Bigl(\frac{\mws}{s} \Bigr)$ 
& $\simeq \, 1.97 $ & \hskip 0.5 cm
 ${\cal A}^0_{\pm,\pm}({\gamma\gamma\to HH})$ & \qquad\eqref{eq:gghhmm} &&  \\
&&&&&&\\ 
 $|\btwohc|$ &$\qquad\leq \,\frac{16\sthws}{3\,\alpha} \Bigl(\frac{\mws}{s} \Bigr)$
& $\simeq \, 1.01 $ & \hskip 0.5 cm
$ {\cal A}^0_{00}({WW\to HH}) $& \qquad\eqref{eq:wwhh00} && \\ 
& & & \hskip 0.5 cm $ {\cal A}^0_{00}({ZZ\to HH}) $ &\qquad\eqref{eq:zzhh00} &&\\
\hline
 \hline
\end{tabular}
\caption{\small\em{ Left panel of the table exhibit the   unitarity constraints on the 11 linearly
 independent anomalous couplings and their corresponding partial wave amplitudes as given in appendix \ref{subsec:helamp}.    These bounds are obtained by retaining  only one non-zero anomalous coupling at a time for all the processes analysed in this article. Right panel of the table show the stringent upper bound of anomalous couplings which are linearly dependent on  $\dgoz$ as given in equation \eqref{eq:cond-final}.  } }
\label{tab:ind-bounds}
\end{table}
\end{center}
\end{widetext}
\vskip -10 cm

 \par In this article, we compute   the partial wave helicity amplitudes for all vector boson scattering processes $V_1\, V_2 \to V_3\, V_4 $ and  also for all processes where  one or more  vector bosons $V_i$ are replaced by the scalar Higgs and their corresponding helicity $\lambda_i$ by zero as mentioned in Section \ref{sec:intro} (with the choice of
momenta and polarisations listed in Appendix ~\ref{app:notation}). We further investigate the high energy behaviour of these amplitudes  as a function of the 23 anomalous couplings (five TGC, six QGC, four $VVH$, four $VVHH$, two $H^3$ and two $H^4$) by expanding them in powers of
energy {\it viz.} $s/m^2$. 

\par The  partial wave amplitudes for the processes considered by us grow with  energy ($\sqrt{s}$) as either  \orderss or \ordersthalf.  The unitarity condition given in equation \eqref{eq:unitarity-cond} would then impose
\bea
c_1 \left(\frac{s}{m^2}\right)^{1/2} + c_2 \left(\frac{s}{m^2}\right) + 
c_3 \left(\frac{s}{m^2}\right)^{3/2} +c_4 \left(\frac{s}{m^2}\right)^2 \leq
\frac{1}{2}  \, ,\nn \\
\label{eq:unitarity-cond1}
\eea
where each of $c_i$ is a  linear combination of the anomalous couplings.  Working along  the perturbative unitarization of gauge dynamics in the SM (as mentioned in Section \ref{sec:intro}) and  beyond the SM scenarios (for example in reference ~\cite{Csaki:2003dt}), we expect that the cancellation of $s^2$  and $s^{3/2}$ terms in all the $ 2\to 2$ scattering amplitudes can be realized among the guage  mediated diagrams even in the presence of anomalous couplings   involving the Higgs boson  and gauge bosons. This provides us a clue as well as a conservative choice for  satisfying equation \eqref{eq:unitarity-cond1}. Hence, we enforce this choice by demanding the linear combinations $c_3$ and $c_4$ to be zero. The resulting relations among the anomalous couplings are then exploited along with the equation \eqref{eq:cond-gaugeinv3} to reduce number of independent parameters and obtain
\begin{eqnarray}
 \lz & = &\lgam =\dkg = 0 \label{eq:cond-dkg}\\
2 \dgoz &=& \dczeroz = \dcoz = 2\dczerozg = 2\dkz \nn\\
&=& \frac{\dczerow}{\cos^2{\theta_W}} = \frac{\dcow}{\cos^2{\theta_W}} 
\label{eq:cond-final}
\end{eqnarray}
\par Thus,  of the above ten anomalous couplings, three vanish and rest seven are related among themselves leaving us with only one  independent coupling which we take to be \dgoz. We are now left with a set of following fourteen linearly independent anomalous couplings:
\bea
&& \dgoz, \, \dcozg, \,\atwog,\, \atwozg,\, \atwoz, \,  \atwow ,\nn\\
&& \atwogh,\, \atwozgh,\, \atwozh, \,  \atwowh ,\nn\\
&&  \dbohc,\, \dbohq,\, \btwohc \, {\rm and } \,\,\btwohq. \label{indepano}
\eea
\subsection{Unitarity Bound}
\noindent 
After using the relations given by \eqref{eq:cond-dkg}-\eqref{eq:cond-final}, the most divergent partial wave helicity amplitudes at high energies of all the  scattering processes considered by us are at most either of \orders or \ordershalf and are listed in equations \eqref{eq:zzzz-00mm}-\eqref{eq:gghhmm} of Appendix ~\ref{subsec:helamp}. Note that the higher partial wave amplitudes ${\cal A}^J$ with $J>0$ 
 grow with  energy slowly compared to ${\cal A}^0$ and thus give less stringent
 bounds on the couplings. Hence only the lowest partial wave amplitudes ${\cal A}^0$
 are listed  in the appendix. Higher partial scattering amplitudes $( J>0)$ are listed only for the
 cases where they provide independent bounds on the anomalous couplings.

\par However, the quartic Higgs boson self couplings  \btwohq which appears only in $HH\to HH$ scattering process do not show any bad high energy behaviour and hence it cannot be constrained from the energy dependent unitarity argument given in \eqref{eq:unitarity-cond}.
On the same note  anomalous triple \dbohc  and quartic \dbohq Higgs couplings  which do not contribute to any
 amplitude that grows with energy, cannot be  constrained from perturbative unitarity.

\par  With the  help of relations  \eqref{eq:unitarity-cond}, \eqref{eq:cond-dkg} and \eqref{eq:cond-final} we are now equipped to extract the unitarity constraint $\left\vert {\cal R} e ({\cal A}^J(s)) \right\vert \le 1/2$ either on the individual anomalous couplings or on the linear combination of anomalous couplings  from the remaining all non-zero partial wave amplitudes which are of the  \orders or \ordershalf.    We calculate the absolute upper bound of the anomalous couplings, by considering the effect on the high energy behaviour of the partial wave amplitudes for all the processes simultaneously keeping one anomalous coupling at a time and report the most stringent ones for the independent couplings given in equation~\eqref{indepano} in the left panel of Table~\ref{tab:ind-bounds}.  The bounds on other six dependent anomalous couplings related to the \dgoz via \eqref{eq:cond-final}  may be derived from the obtained constraints and  are given in the right panel of the same Table. While computing the upper bound on anomalous couplings we  have used \cite{pdg}
\begin{align}\label{eq:param-value}
\begin{split}
\alpha^{-1}(\mz) =& 127.916 \, , \\
\sthws(\mz) =& 0.23116 \, , \\
\mz =& 91.1879 \gev .
\end{split}
\end{align}

\par Extending our analysis, we  allow simultaneous variation of two or more non-zero anomalous couplings  and search for a constrained region in the parameter space. Thus, we  consider all  such partial wave amplitudes that depend upon more than one anomalous coupling.

\par We observe that the anomalous couplings \dgoz , \dcozg and \atwozg  affect the partial
amplitudes given by equation \eqref{eq:wwzg00mm}: ${\cal A}^0_{00\pm\pm}({WW\to Z\gamma})$.
The constrained parameter region obtained from the perturbative unitarization of these partial wave amplitudes  along with the constraints from ${\cal A}^0_{0000}({WW\to ZZ})$ and
$ {\cal A}^0_{00\pm\pm}({ZZ \to Z\gamma})$ (as listed in Table~\ref{tab:ind-bounds}) is given by
\bea\footnotesize
- 1 -\frac{\alpha s} {2\tthw\mws} \dgoz  \leq  f \leq 
 1 - \frac{\alpha s} {2\tthw\mws} \dgoz;&&\nn\\
 {\text{with}} \quad f =\frac{\alpha} {2\tthw}\frac{s}{\mws}\left[\frac{1}{4\tthw} \atwozg-\dcozg \right].&& 
\label{eq:bound2}
\eea
This region is displayed in Figure~\ref{fig:bound2}.
Similarly, the bounds on \dgoz and \atwow are not independent as they are also 
related by various amplitudes for the process $WW \to WW$ given in equation
 \eqref{eq:wwwwm000}. The region on the ($\atwow-\dgoz$) plane simultaneously allowed by the non-violation of unitarity of these amplitudes and the constraints from  amplitudes 
 ${\cal A}^0_{0000}({WW\to ZZ})$, $ {\cal A}^0_{\pm\pm 00}({WW\to WW})$, 
 and $ {\cal A}^0_{\pm\pm 00}({WW\to ZZ})$ 
 is given by
\bea \footnotesize
&& -\,C_1 + 2(4-3\sec^2{\theta_W}) \Bigl(\frac{\rts}{\mw} \dgoz\Bigr)
 \leq \Bigl(\frac{\rts}{\mw} \atwow\Bigr)
\nn\\
&& \leq C_1 + 2(4 - 3\sec^2{\theta_W}) \Bigl(\frac{\rts}{\mw} \dgoz\Bigr)
\nn\\
&&{\rm with} \quad C_1 = \frac{32\sqrt{2} \tthws}{\pi \alpha }.
\label{eq:bound3}
\eea
This constrained parameter space is plotted in  Figure~\ref{fig:bound3}.
\section{Dimension Six Operators}
\label{sec:d6op}
As discussed in  Section~\ref{sec:intro},  model independent NP effects  can also be investigated by adding gauge invariant higher dimensional operators to the SM Lagrangian. The present precision of the data allows us to parametrize the deviations 
of the SM couplings in terms of coefficients of these higher dimension operators. We consider  operators upto dimension six for our analysis {\it i.e.} upto the order of $1/\Lambda^2$ in the expansion given in equation \eqref{fulllagrangianop6}. A complete list
of such operators is listed in the classic paper of reference \cite{Buchmuller:1985jz} and
are classified again in reference \cite{Grzadkowski:2010es}. 
\par Restricting ourselves to the CP-even dimension six operators which are 
 relevant for the scattering processes considered in this article, {\it i.e} the
 ones that modify the Higgs and electroweak gauge bosons couplings, we get the following ten operators:
\bea
{\cal O}_{WWW} &=& \,\mbox{Tr}[{\hat W}_{\mu\nu}{\hat W}^{\nu\rho}{\hat W}_{\rho}^{\mu}],\nn
\\
{\cal O}_W &=&\, (D_\mu\Phi)^\dagger {\hat W}^{\mu\nu}(D_\nu\Phi),\nn
\\
{\cal O}_B &=&\, (D_\mu\Phi)^\dagger {\hat B}^{\mu\nu}(D_\nu\Phi),\nn
\\
{\cal O}_{BB} &=&\, \Phi^{\dagger} {\hat B}_{\mu \nu} {\hat B}^{\mu \nu} \Phi,\nn
\\
{\cal O}_{WW} &=& \,\Phi^{\dagger} {\hat W}_{\mu \nu} {\hat W}^{\mu \nu} \Phi ,\nn\\
{\cal O}_{BW} &= & \,\Phi^{\dagger} {\hat W}^{\mu \nu} \Phi {\hat B}_{\mu \nu}, \nn\\
{\cal O}_{\Phi,1} &=& \, \left((D_\mu \Phi^\dagger )\Phi \right) \left(\Phi^\dagger D^\mu \Phi\right),\nn\\
{\cal O}_{\Phi,2} &=& \,\frac{1}{2} \partial_\mu \left(\Phi^{\dagger} \Phi \right)
\partial^\mu \left(\Phi^{\dagger} \Phi \right),\nn
 \\
{\cal O}_{\Phi,3} &=&\, -\frac{1}{3}\left(\Phi^{\dagger} \Phi \right)^3,\nn\\
{\cal O}_{\Phi,4}&=& \,(D_\mu \Phi)^\dagger (D^\mu \Phi ) \Phi^\dagger\Phi.
\label{eq:op-list}
\eea
 Here $\Phi$ is the Higgs doublet represented in the unitary gauge as 
\begin{equation}
\Phi=\frac{1}{\sqrt{2}}\begin{pmatrix}0\\v+h(x)\end{pmatrix}.
\end{equation}
The covariant derivative  along with  field strength tensors ${\hat W}^{\mu \nu}$ and ${\hat B}^{\mu \nu}$ are defined as
\vskip -0.5cm
\begin{align}
\begin{split}
D_\mu & = \partial_\mu + \frac{i}{2} g \tau^\text{\tiny \em I} W^\text{\tiny \em I}_\mu + \frac{i}{2} g' B_\mu \,,
\\
{\hat B}_{\mu \nu} & = \frac{i}{2} g' (\partial_\mu B_\nu - \partial_\nu B_\mu)\, ,
\\
{\rm and} \ {\hat W}_{\mu\nu} & = \frac{i}{2} g\tau^\text{\tiny \em I} (\partial_\mu W^\text{\tiny \em I}_\nu
    - \partial_\nu W^\text{\tiny \em I}_\mu + g \epsilon_\text{\tiny \em IJK} W^J_\mu W^K_\nu ).
\end{split}
\end{align}
\noindent 
It may be noted that, of the ten operators listed in equation~\eqref{eq:op-list}, 
only one operator, namely, 
 ${\cal O}_{\Phi,3} $ gives an additional  contribution to the 
scalar Higgs boson potential  and hence modifies the  minima of the SM potential. This, in turn, modifies the SM vacuum expectation value  $v_{\text{\tiny SM}}^2=-\mu^2/\lambda$ to
\bea
 \frac{v^2}{2}
 &\simeq& \left( \frac{v_{\text{\tiny SM}}^2}{2}\right)\left[ 1 -\left(\frac{ \fphith}{4\Lambda^2} \right)v_{\text{\tiny SM}}^2  \right].  
 \label{eq:newvev} 
 \eea
Further, inclusion of the operators ${\cal O}_{\Phi,1} $,
 ${\cal O}_{\Phi,2} $,  and ${\cal O}_{\Phi,4} $ modifies the kinetic term of the
Higgs field, leading to a redefinition of the Higgs boson field and the Higgs boson mass $m_H $  \cite{Corbett:2014ora} as  
\begin{eqnarray}
&&H\simeq \left[1+\frac{v^2}{4\Lambda^2}(f_{\Phi,1}+2f_{\Phi,2}+f_{\Phi,4}) \right]\, h; \label{higgsrenorm}\\
&&m_H^2\simeq 2\lambda v^2\left[1-\frac{v^2}{2\Lambda^2}
\left(f_{\Phi,1}+2f_{\Phi,2}+f_{\Phi,4}+\frac{f_{\Phi,3}}{\lambda}\right)\right]. \nonumber\\
\label{higgsmassrenorm}
\end{eqnarray}
\par The operators ${\cal O}_{BW}  $  and ${\cal O}_{\Phi,1} $ have a tree level
 effect on precision electroweak observables and therefore are subject to very
 strict constraints \cite{Dutta:2008bh}. Hence we do not constrain these operators in our analysis. 


\subsection{Relation to Anomalous Couplings}
A given dimension six operator can be expanded  in terms of a set of independent Lorentz structures appearing with the same coefficient $f_i$. 
On comparing with   the effective Lagrangian given in equation \eqref{lag:pheno}, we can express  the anomalous couplings as a linear combination of the coefficients $f_i$  (see reference \cite{Hagiwara:1996kf} and \cite{Dutta:2008bh}). Below we give relations of TGC and QGC  with the coefficients of the above mentioned dimension six operators:
\begin{eqnarray}
\lambda_\gamma &=& \lambda_Z = f_{WWW}\frac{3g^2m_W^2}{2\Lambda^2} \, , \label{eq:tgcfromEFT1}
\\
\dkz &=& (f_Wu-f_B\tan^2\theta_W)\frac{m_W^2}{2\Lambda^2}\,, \label{eq:tgcfromEFT2a}
\\
\dkg &=& (f_W+f_B)\frac{\mws}{2\Lambda^2},\,\,\,\, {\rm and} \label{eq:tgcfromEFT2}
\\
\dczeroz &=&\dcoz = 2 \dczerozg = 2 \dcozg= \frac{\dczerow}{\cthws} = \frac{\dcow}{\cthws}
\nn\\
&=& -2 \dgoz =  -f_W\frac{\mzsq}{\Lambda^2}.  \label{eq:tgcfromEFT3}
\end{eqnarray}
\noindent 
\par Based on the analysis performed in previous section we attempt to constrain the coefficients of dimension six operators. 
Combining the relations \eqref{eq:tgcfromEFT3} 
 with the unitarity constraints from  \eqref{eq:cond-final} we get  
\bea
\dgoz=\Delta c^\text{\tiny $VV^\prime$}_{_i}=0. \label{opunirel0}
\eea
Further using \eqref{eq:cond-dkg}, \eqref{eq:tgcfromEFT1},  \eqref{eq:tgcfromEFT2a} and \eqref{eq:tgcfromEFT2}, we get $f_{WWW}=f_W = f_B =0$.
\par We now parameterize the coefficients of remaining five operators in terms of five dimensionless  parameters defined as
\bea
&&d_{2} = \frac{\mws}{\Lambda^2} f_{\Phi,2}; \quad d_{3} \,=\, \frac{\mws}{\Lambda^2} f_{\Phi,3};\,,\quad d_{4} \,=\, \frac{\mws}{\Lambda^2} f_{\Phi,4}; \nonumber\\
&& d = - \frac{\mws}{\Lambda^2} f_{WW} \quad {\rm and} \quad  d_B \,=\, - \frac{\mws}{\Lambda^2} \tthws f_{BB}.
\label{eq:hvvfromEFT}
\eea
\noindent We re-write all the  anomalous  $VVH$ and $VVHH$ Higgs-gauge bosons couplings  and Higgs  boson self interactions
 in terms of these   dimensionless  parameters $d_i$'s and TGC (see 
refs.~\cite{Dutta:2008bh,Corbett:2012dm,Achard:2004kn,Barger:2003rs}) \footnote{Note that we do not take into account the operators  ${\cal O}_{\Phi,1}$ and  ${\cal O}_{BW} $ as mentioned earlier}. 
\bea
\daow &=& \daoz \,=\, \frac{\sthws}{4\pi\alpha}\left(3 d_4 -2 d_2 \right),
 \label{eq:daow}\\
\daowh &=& \daozh \,=\, \frac{\sthws}{4\pi\alpha}\left(5 d_4 -2 d_2 \right),
 \label{eq:daoz}\\ 
\atwow &=& \atwowh \,=\, 2\secthws \Bigl[d + \cthws \,\dgoz\Bigr],
\label{eq:atwow}\\
\atwoz &=&  \atwozh \,=\, 2\,\secthws\Bigl[d\cthws + d_B \sthws
\nn\\
 && \hspace{1cm} +\,\dgoz \cos{2\theta_W} + \dkg \tthws\Bigr],
\label{eq:atwoz}\\
\atwozg &=& \atwozgh \,=\, 2\tthw \Bigl[d-d_B + \dgoz -\frac{\dkg}{2\cthws}\Bigr],
\nn\\
\label{eq:atwozg}\\ 
\atwog &=& \atwogh \,=\, 2\,\secthws\Bigl[d \sthws+ d_B\cthws \Bigr].  \label{eq:atwog}
\eea
 Using \eqref{eq:cond-gaugeinv1} and \eqref{eq:cond-gaugeinv1a},  along with \eqref{eq:daow}, \eqref{eq:daoz} we get
\bea
d_2 =d_4 =0.\label{opunirel1}
\eea
Anomalous couplings inducing Higgs boson self interactions $H^3$ and $H^4$  are related to the parameters $d_i$'s  as
\bea
\dbohc &=& \frac{\sthws}{12\pi\alpha} \left[ -6d_2 -3d_4 +\frac{8v^2}{\mhsq} d_3\right], \label{eq:dbohc}\\
\dbohq &=&\frac{\sthws}{2\pi\alpha} \left[ -2d_2 -d_4 +\frac{8v^2}{\mhsq} d_3\right],\label{eq:dbohq}\\
\btwohc &=& \btwohq=\frac{\sthws}{3\pi\alpha} \left[ 2d_2 +d_4 \right].\label{eq:btwohcq} 
\eea
\begin{figure}[!ht]
  \centering 
 \includegraphics[width=0.47\textwidth]{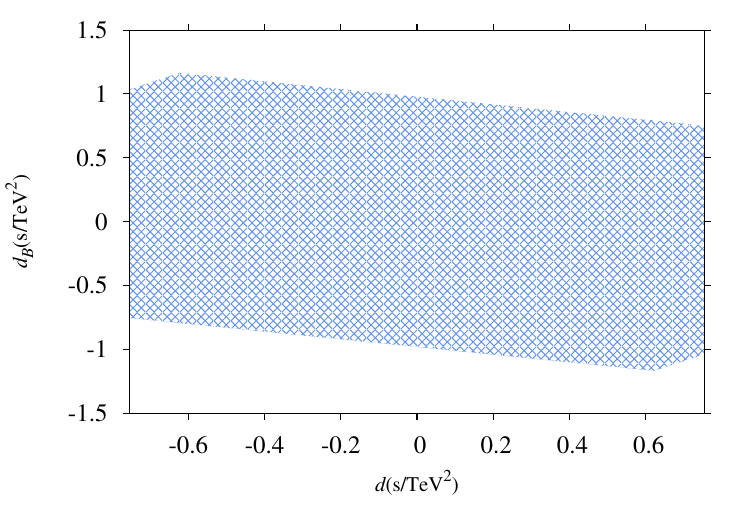}  
\caption{\small\em The shaded enclosed region on the $(d-d_B)-$ plane corresponds to    unitarity constraints from partial wave amplitudes  as given in equations  \eqref{ddbcons0}-\eqref{ddbcons3}}
\label{fig:bound-d-db}
\end{figure}
However, the constraint given in equation  \eqref{opunirel1} guarantees the vanishing of the anomalous couplings  $\btwohc$ and $\btwohq$. As a consequence \dbohc and \dbohq depend only on  the  dimensionless parameter $d_3$.  
\begin{figure}[!ht]
  \centering
\includegraphics[width=0.47\textwidth]{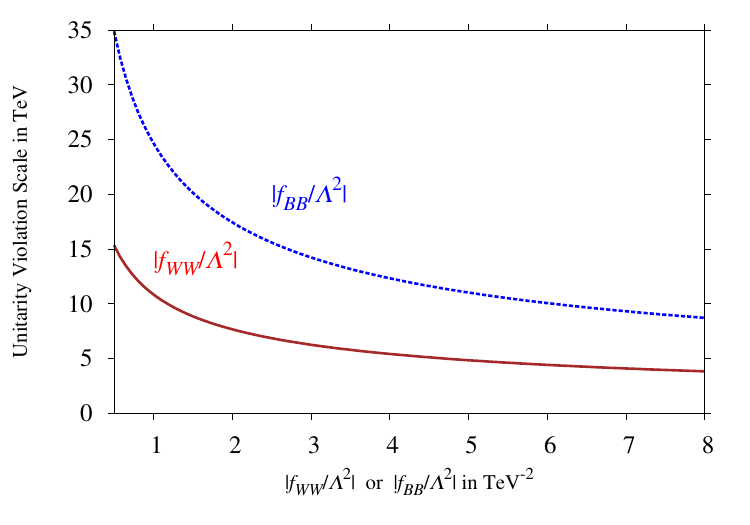}
\caption{\small\em Variation of unitarity violation energy with coefficients  $f_{WW}/\Lambda^2$ and $f_{BB}/\Lambda^2$. These coefficients are varied within  limits derived from combined analysis of LHC and Tevatron data at 90 \% CL. }
\label{fig:rts_limit}
\end{figure}
\par Thus operator analysis has reduced the number of linearly independent parameters to three, namely $d_3$, $d$ and $d_B$, which are essentially  the dimensionless  coefficients of operators ${\cal O}_{\Phi,3}$, ${\cal O}_{WW}$ and ${\cal O}_{BB}$ respectively. This  is  in contrast to fourteen linearly independent anomalous couplings based on the partial wave analysis of the Lorentz structures given  in equation \eqref{indepano} of previous Section. Out of these three, non-violation of perturbative unitarity constraints only $d$ and $d_B$ while  $d_3$ remains unconstrained as it does not appear as a coefficient of \orders or \ordershalf  terms in any of the partial wave amplitudes.

\par With the vanishing of anomalous TGC and QGC \eqref{opunirel0}, the dimensionless coefficients $d$ and $d_B$  are related to four $VVH$ (or $VVHH$) couplings, taking any two  at a time. We depict the allowed region constrained by unitarity of  all partial wave amplitudes on $(d,\,d_B)-$plane in Figure  \ref{fig:bound-d-db}. The enclosed  region is constrained by following four inequalities arising   from \eqref{eq:wwzzmm00}, \eqref{eq:wwwwmm00} \& \eqref{eq:wwhhmm}, \eqref{eq:wwzg00mm} \& \eqref{eq:zghhmm}, 
\eqref{eq:wwgg00mm} \& \eqref{eq:gghhmm} and \eqref{eq:wwzz-00mm} \& \eqref{eq:zzhhmm} respectively
\bea
&\left\vert d\right\vert &\le  \frac{4\sin^2\theta_W}{\alpha } \left(\frac{\mws}{s}\right),\label{ddbcons0}\\
&\left \vert d_B -d\right\vert &\le \frac{4\tthw}{\alpha } \left(\frac{\mws}{s}
\right) ,
\label{ddbcons1}\\
&\left \vert d_B \cot^2\theta_W+ d\right\vert &\le \frac{4}{\alpha }
 \left(\frac{\mws}{s}\right) \, ,
\label{ddbcons2}\\
&\left \vert d_B + d\cot^2\theta_W\right\vert& \le \frac{4\mws}{\alpha s}. \label{ddbcons3}
\eea 
\noindent Unitarity bounds on anomalous $VVH$ and $VVHH$ couplings can thus be translated to $d$ and $d_B$ from the boundary of the enclosed shaded region of Figure~\ref{fig:bound-d-db}. Translating in terms of the coefficients of the dimension six operators, we get the most stringent bounds  from \eqref{ddbcons1} and \eqref{ddbcons2} to be
\bea
 \left\vert\frac{f_{WW}}{\Lambda^2}\right\vert \le\frac{4\sthws}{ \alpha}\Bigl(\frac{1}{s}\Bigr) &=& \left(\frac{118}{s}\right) \quad {\rm and} \label{eq:bound-fww2}\\
\left\vert \frac{f_{BB}}{\Lambda^2}\right\vert \,\le \, \frac{4(1+\tthw)}{ \secthws \,\alpha\, s}& =&\left(\frac{609}{s}\right).
\label{eq:bound-fbb3}
\eea
\par However, keeping only one coupling at a time, the most stringent unitarity bound on $f_{WW}/\Lambda^2$ remains same as given in equation \eqref{eq:bound-fww2} while upper bound on $f_{BB}/\Lambda^2$ is further lowered and is given as
\bea
\left\vert\frac{f_{BB}}{\Lambda^2}\right\vert \le \frac{4\mws}{\alpha s} = \left(\frac{512}{s}\right).
\label{eq:bound-d1}
\eea
\section{Constraints from Experiments}
In this section we discuss  the experimental constraints on anomalous couplings.
Adhering to the conditions given in equations \eqref{eq:cond-gaugeinv2} and \eqref{eq:cond-gaugeinv3}, the existing LEP limit on TGC  \cite{Schael:2013ita} along with recent data from LHC \cite{Chatrchyan:2013yaa} are summarized in the Table \ref{tgcexplim}.
\begin{table}[h!]
\begin{tabular}{ccc}\hline\hline
Anomalous TGC & \,\,\, LEP\,\,\,& \,\,\,LHC\,\,\,\\
\hline
$\left\vert\Delta g^\text{\em\tiny Z}_{_1}\right\vert$ & 0.020& 0.095\\
$\left\vert\lz \right\vert $ &0.022 & 0.048\\
$\left\vert\Delta \kappa_\gamma \right\vert$  & .042& 0.22\\
\hline\hline
\end{tabular}
\caption{{\em Experimental Limits on anomalous Triple gauge boson couplings assuming the custodial $SU(2)$ symmetry. }}
\label{tgcexplim}
\end{table} 
\noindent However,  one can obtain much less stringent bound on these couplings by relaxing the  custodial and gauge symmetry and a similar analysis have been performed with LHC data \cite{ATLAS:2012mec} to give
\begin{align}
 - 0.135 \leq &\Delta\kappa_\gamma \leq 0.190, \nn\\
 - 0.373 \leq &\Delta g_1^Z \leq 0.562, \nn\\
 - 0.078 \leq &\Delta\kappa_Z \leq 0.092, \nn\\
 - 0.152 \leq &\lambda_\gamma \leq 0.146, \nn\\
 - 0.074 \leq &\lambda_Z \leq 0.073.
\end{align}  

\par  LEP bounds on the coefficient of operators ${\cal O}_{BB}$ and ${\cal O}_{WW}$   involving Higgs gauge boson  coupling can be read out from Figure 6 of reference ~\cite{Achard:2004kn} ( for $\mh  = 125 \gev$) and expressed in terms of the upper limit on the magnitude of $\left\vert d_B\right\vert \,\lesssim \,0.05$ and $\left\vert d\right\vert \,\lesssim  \,0.2$.  Using these upper limits on $d_B$ and $d$ we  find that unitarity is not violated  upto 4.8 and 2 ~TeV respectively. 

\par Further, adding LHC data  provides stringent
limits, particularly when the Higgs to two photon decay signal strength is taken into account~\cite{Masso:2012eq}.   In this reference, the authors study the constraints on the dimension six operators by analysing LHC data from Higgs decays  $H\to \gamma\gamma $, $H\to WW$ and $H\to ZZ$ channels. The one parameter bounds on  $\e_{WW}\equiv v^2 f_{WW} /\Lambda^2$ and  $\e_{BB}\equiv v^2 f_{BB} /\Lambda^2$ from ATLAS and CMS data obtained by them for diphoton channel at 95 \% CL  translate into
\begin{align}\label{eq:masso-lhc}
\begin{split}
\frac{f_{WW}}{\Lambda^2} , \frac{f_{BB}}{\Lambda^2} & \in \left[ -3.47, 0.496 \right] {\rm TeV}^{-2},
\\
& {\rm Diphoton\, Channel:\, ATLAS}
\\
\frac{f_{WW}}{\Lambda^2} , \frac{f_{BB}}{\Lambda^2} & \in \left[ -3.80, 0.826 \right] {\rm TeV}^{-2}.
\\
& {\rm Diphoton\, Channel:\, CMS}
\end{split}
\end{align}
Combining the analysis from ATLAS and CMS, we compute the lowest energy scale where unitarity would be violated in the presence of these dimension six operators. Taking one operator at a time, the unitarity violation scale becomes $\sim 6\tev$ and $13\tev$ respectively for $f_{WW}/\Lambda^2$ and $f_{BB}/\Lambda^2$. 
We depict the variation of the  unitarity violating  scale with $\left\vert f_{WW}/\Lambda^2\right\vert$ and $\left\vert f_{BB}/\Lambda^2\right\vert$ in Figure \ref{fig:rts_limit}. In this figure, we have considered both these coefficients to vary within the allowed range given by the combined analysis.

\par A global fit to the existing LHC and Tevatron data  has been performed in reference ~\cite{Corbett:2012ja} allowing simultaneous determination of the parameters quantifying the Higgs boson couplings to the electroweak gauge bosons and the other SM particles. Using their best fit value 1.5 (-1.6)$\, {\rm TeV}^{-2}$ for $f_{WW}/\Lambda^2$ ($f_{BB}/\Lambda^2$), we observe that unitarity is not violated upto energies $\sim 9\, (19)\tev$. Further if the operator $f_{WW}/\Lambda^2$ ($f_{BB}/\Lambda^2$) is allowed to be as large as the largest value of the 90\% CL regions which is 8.2 (7.5)$\, {\rm TeV}^{-2}$ , the unitarity is preserved until $\sim 4\, (9)\tev$. 

\section{Conclusions}
\label{sec:conc}
In this article, we have attempted to address perturbative unitarity of the vector boson
scattering processes in the presence of anomalous couplings associated with the
pure gauge sector (TGC and QGC), the Higgs boson - gauge boson sector  ($V_1V_2H$, $V_1V_2HH$) and Higgs boson self interactions. We start with twenty three anomalous couplings involved in $VV$ and/or $HH$
scattering processes, taking all of them to be independent. Our observations are
summarised below:
\begin{enumerate}
\item[(a)] We adopt the correct procedure for perturbative unitarization by analysing all  dominant terms in the helicity amplitudes unlike reference \cite{Corbett:2014ora}. At high energies, the helicity amplitudes corresponding to the gauge
boson scattering in  processes grow as \orderss and/or \ordersthalf$\!$.
On demanding these divergent terms in the amplitudes to vanish identically, we
are left with fourteen independent anomalous couplings given in equation \eqref{indepano}.
\par However,   three anomalous couplings  which are related to Higgs boson self interactions, do not generate  such terms in helicity amplitudes that grow with  energy  and hence they  could not be  constrained from the perturbative unitarity arguments.
\par On unitarizing all non-zero helicity amplitudes which
grow as  either \orders or \ordershalf, we can successfully constrain remaining eleven anomalous couplings.  In Figures \ref{fig:bound2}, and \ref{fig:bound3} we  plot the constrained regions of the linear combination of two anomalous couplings.    
\par Upper limits on each of these couplings are computed taking one coupling to be operative  at a time. The upper bounds of the independent and dependent anomalous couplings  are presented in the Table \ref{tab:ind-bounds} in units of $(1\, \tev^2/s)$. Inverting the argument, the
 Table can also be used to read off the energy scale upto which unitarity is not
violated for a given value of coupling. Using the current  LEP bound on \dgoz~\cite{Schael:2013ita}, $\left\vert\dgoz\right\vert\le 0.016$ and reading out the constraint from the right column of the Table, we find that perturbative unitarity is not violated upto $\rts \sim 7$~TeV.

\item[(b)] Assuming that the contribution to the anomalous couplings are restricted to have arisen from five CP-even dimension six operators, we find that the perturbative unitarity requires vanishing of all anomalous TGCs and QGCs. Study of the $VV-$ scattering processes shows that the SM along with anomalous couplings in the Higgs - gauge boson sector can preserve unitarity at least upto $\simeq 4\tev$ with 90 \% CL. This result is also in agreement with the recent work in reference \cite{Choudhury:2012tk}, where the authors have considered the anomalous Higgs coupling with the top quark and using the best fits of preliminary LHC Higgs data they show that unitarity can be preserved upto 4 TeV unless NP takes over. 
\par Using the best fit values of the combined analysis with Tevatron and LHC data \cite{Corbett:2012ja}, we observe that the unitarity validation scale can be raised upto 9 TeV.
\item[(c)] Comparing our results with that of reference \cite{Corbett:2014ora}, we observe that, unlike theirs, we have only two linearly independent dimension six operators which fix the perturbative unitarity violation scale. In addition we provide the limits on all anomalous TGC, QGC, $VVH$, $VVHH$ and Higgs boson self couplings.  
\end{enumerate}
\vskip 0.5 cm
With more data from CMS and ATLAS at LHC, we expect to improve the unitarity bound on the anomalous couplings. Accordingly, the unitarity violation scale can be raised with the shrinking of the
allowed region in anomalous couplings. On the contrary, if we find these anomalous couplings to be rather large than one needs to invoke a careful study of divergence cancellations with the inclusion of new physics spectrum, to respect unitarity. 

\acknowledgments

The authors  thank  Sudhendu Rai Choudhury  and Debajyoti Choudhury for fruitful discussions which helped us to bring out an improved version of our earlier work. MD and SD acknowledge the partial financial support from the CSIR grant No. 03(1340)/15/EMR-II and the DST grant No. SR/S2/HEP-12/2006.  RI acknowledges the DST-SERB  grant No. SRlS2/HEP-13/2012 for the partial
financial support. MD and SD would like to thank IUCAA, Pune for  the hospitality where part of this
work was completed.

\begin{appendix}
\setcounter{equation}{0}
\section{Notations and Conventions of Momenta and Polarizations}
\label{app:notation}

In our present article we study all $2 \to 2$ gauge boson scattering
processes. Here we define the choice of
momenta and polarizations vectors. 

\par  For all our scattering process the masses of initial particles are identical. Therefore, in CM reference
frame, the momenta and polarization of initial particles are given as
\begin{align}
\begin{split}
 k_1 \equiv& \frac{\rts}{2} \left( 1,0,0, \b_{_V} \right); \\
 k_2 \equiv& \frac{\rts}{2} \left( 1,0,0,-\b_{_V} \right),
\end{split}
  \\
\begin{split}
 \e^\pm(k_1)\equiv& \frac{1}{\sqrt2} \left( 0,\pm1   ,- i,0       \right); \\
 \e^\pm(k_2)\equiv& \frac{1}{\sqrt2} \left( 0,\mp1   ,- i,0       \right); \\
 \e^0(k_1) \equiv& \frac{\rts}{2 m_{_V}} \left( \b_{_V},0,0, 1 \right) \\
 \e^0(k_2)\equiv& \frac{\rts}{2 m_{_V}} \left( \b_{_V},0,0,-1 \right).
\end{split}
\end{align}
where $\b_{_V} = \sqrt{1 - 4 m^2_{_V}/s}$, $\rts$ being the CM energy
and $m_{_V}$ the mass of the corresponding gauge boson (here $W$ or $Z$). 

\par Similarly, for the processes $WW \to \gamma \gamma$, $WW \to WW$, $WW \to ZZ$,  $ZZ \to ZZ$ the momenta and transverse polarization of final particles are defined as 
\begin{align}
\begin{split}
 k_3\equiv& \frac{\rts}{2}
       \left( 1, \b_{_{V^\prime}}\sth,0, \b_{_{V^\prime}}\cth \right); \\
 k_4\equiv& \frac{\rts}{2}
       \left( 1,-\b_{_{V^\prime}}\sth,0,-\b_{_{V^\prime}}\cth \right)
\end{split} \\
\begin{split}\label{trans_pol}
 \e^\pm(k_3) \equiv& \frac{1}{\sqrt2} \left( 0,\pm\cth, -i,\mp\sth \right); \\
 \e^\pm(k_4) \equiv& \frac{1}{\sqrt2} \left( 0,\mp\cth, -i,\pm\sth \right).
\end{split}
\end{align}
where $V^\prime$ implies $W,Z$ or $\gamma$, $\cth = \ct$ and
$\sth = \st$, $\theta$ being scattering angle which the angle
between ${\bf k}_1$ and ${\bf k}_3$.
\noindent Except for the photons all other final state  massive gauge bosons have longitudinal polarisation which is defined as
\begin{align}
\begin{split}
 \e^0(k_3) \equiv& \frac{\rts}{2 m_{_{V^\prime}}} \left( \b_{_{V^\prime}}, \sth,0, \cth \right); \\
 \e^0(k_4) \equiv& \frac{\rts}{2 m_{_{V^\prime}}} \left( \b_{_{V^\prime}},-\sth,0,-\cth \right).
\end{split}
\end{align}

\par The process $WW \to Z \gamma$ is unique as here the final state consist of  particles with unequal masses. We define the momenta of the final state particles for $WW \to Z \gamma$ as 
\begin{align}
\begin{split}
k_3 \equiv& \frac{1}{2 \rts}
       \bigg( (s+\mz^2), \\
       & (s-\mz^2)\sth,0, (s-\mz^2)\cth \bigg); \\
k_4 \equiv& \frac{1}{2 \rts}
       \bigg( (s-\mz^2), \\
       &-(s-\mz^2)\sth,0,-(s-\mz^2)\cth \bigg).
\end{split}
\end{align}
\noindent The transverse polarization of the final state particles $WW \to Z \gamma$ are same as those given in equation \eqref{trans_pol}, while the longitudinal polarization of the $Z(k_3)$ boson is given as
\begin{align} 
\e^0(k_3) \equiv& \frac{1}{2 \rts \mz}
       \Big( (s-\mz^2), \nn\\
       & (s+\mz^2)\sth,0,(s+\mz^2)\cth \Big).
\end{align}
\section{Partial Wave Amplitudes}
\setcounter{equation}{0}
\label{subsec:helamp} 
 After using all conditions and constraints discussed in 
Sections~\ref{sec:formalism} (equations~\eqref{eq:cond-gaugeinv1}--\eqref{eq:cond-gaugeinv1a}) and \ref{sec:partial} (equations~\eqref{eq:cond-dkg}--\eqref{eq:cond-final}), we are left with the following non-vanishing leading partial wave 
amplitudes of the processes we have considered in the high energy limit. 
 Keeping only terms linear in anomalous couplings, we list below the terms 
 of \orders and \ordershalf\! of these partial wave amplitudes ${\cal A}^J_{\lambda_a\lambda_b\lambda_c\lambda_d}$. Note that the $J=0$ 
partial wave amplitudes provide the most stringent unitarity bounds. Hence we provide 
only the amplitudes we have used to calculate the most conservative unitarity bounds on anomalous couplings.\footnote{${\cal A}^0_{\pm\pm00}$ means ${\cal A}^0_{++00} ={\cal A}^0_{--00}$}

\ben
\item \underline{$ZZ\to ZZ$}
\begin{eqnarray}
{\cal A}^0_{\pm\pm00}({ZZ\to ZZ}) &=&{\cal A}^0_{00\pm\pm}({ZZ\to ZZ}))=\nonumber\\
&=&\frac{\alpha\, (s/\mws)}{ 16 \tthws}  \,\atwoz \label{eq:zzzz-00mm}
\eea
\item \underline{$ZZ\to \gamma\gamma$}
\bea
{\cal A}^0_{00\pm\pm}({ZZ\to \gamma\gamma})&=&
=\frac{\alpha\, (s/\mws)}{16\tthws}  \,\atwog \label{eq:zzgg-00mm}
\eea
\item \underline{$ZZ\to Z\gamma$}
\bea
 {\cal A}^0_{00\pm\pm}({ZZ \to Z\gamma})&=& 
  \frac{\alpha\, (s/\mws)}{ 16 \tthws}  \,\atwozg 
 \label{eq:zzzg-m00m}
 \eea
\item \underline{$\gamma\gamma\to \gamma\gamma$}
\vskip 0.05cm
This process takes place with Higgs exchange and since the $\gamma\gamma H$ coupling does not exist at tree level in SM, the helicity amplitudes will be all proportional to square of anomalous coupling $\atwog$. Thus there is no term that is linear in anomalous coupling and this process does not give any constraints at leading order.
\item \underline{$\gamma\gamma\to Z\gamma$}
\vskip 0.05cm
Similar to $\gamma\gamma\to \gamma\gamma$, the helicity amplitudes of this process also depend upon the anomalous coupling \atwozg but all amplitudes are zero if only terms linear in coupling are retained.
\item \underline{$W^+W^-\to \gamma\gamma$}
\bea
&&{\cal A}^0_{00\pm\pm}({WW\to \gamma\gamma})= 
\frac{-  \alpha \, (s/\mws)}{16  \tthws}\,\, \atwog \label{eq:wwgg00mm}
\eea
\item \underline{$W^+W^-\to Z\gamma$}
\bea
&&{\cal A}^0_{00\pm\pm}({WW\to Z\gamma})= 
\frac{\alpha\,(s/\mws)}{16 \tthws}\,\,\times\nonumber\\
&& \Bigl[-\, \atwozg +4 \tthw(\dcozg -\dgoz)\Bigr]  \label{eq:wwzg00mm}
\eea
\item \underline{$W^+W^-\to ZZ$}
\bea
{\cal A}^0_{\pm\pm00}({WW\to ZZ})&=&\frac{-\alpha\, (s/\mws)}{16 \,\tthws}\,\,\atwow \label{eq:wwzzmm00}\\
{\cal A}^0_{00\pm\pm}({WW\to ZZ})&=&\frac{-\alpha\, (s/\mws)}{ 16 \tthws}  \,\atwoz\label{eq:wwzz-00mm}\\
{\cal A}^0_{0000}({WW\to ZZ})&=&\frac{- \alpha\,( s/\mws)}{8\,\sthws}\,\, \dgoz \label{eq:wwzz0000}
\eea
\item \underline{$W^+W^- \to W^+W^-$}
\bea
{\cal A}^0_{00\pm\pm}({WW\to WW}) &=& {\cal A}^0_{\pm\pm 00}({WW\to WW})\nonumber\\
&=&\frac{-\alpha\, (s/\mws)}{16 \,\tthws}\,\,\atwow \label{eq:wwwwmm00}\\
{\cal A}^0_{0000}({WW\to WW})&=&\frac{\alpha\,(s/\mws)}{16\, \sthws}\,\,(4 \sthws-1)\,\,\dgoz \nn\\ \label{eq:wwww0000}\eea
\bea
&&{\cal A}^1_{\mp000}({WW\to WW})=-{\cal A}^1_{0\mp00}({WW\to WW})=\nonumber\\
&&{\cal A}^1_{00\mp0}({WW\to WW})=-{\cal A}^1_{000\mp}({WW\to WW})\nn\\
&&=\frac{\alpha \,\, (\rts/\mw)}{48\,\, \sthws} \,\Bigl[(3-4\cthws)2\dgoz +\cthws\,\atwow  \Bigr] \nn\\ 
\label{eq:wwwwm000}
\eea
\item \underline{$W^+W^- \to HH$}
\bea
 {\cal A}^0_{\pm,\pm}({WW\to HH}) & =& \frac{-\alpha\, (s/\mws)}{16 \,\tthws}\,\,\atwowh    \label{eq:wwhhmm} \\
 {\cal A}^0_{00}({WW\to HH}) & =&  
  \frac{-3\alpha\left( s/(\mws\right)}{32 \sthws} \, \btwohc 
   \label{eq:wwhh00}
\end{eqnarray}
\item \underline{$ZZ \to HH$}
\bea
 {\cal A}^0_{\pm,\pm}({ZZ\to HH})&=& \frac{-\alpha\, (s/\mws)}{16 \,\tthws}\,\,\atwozh      \label{eq:zzhhmm} \\
{\cal A}^0_{00}({ZZ\to HH})& =&
  \frac{-3\alpha\left( s/(\mws\right)}{32 \sthws} \, \btwohc   \label{eq:zzhh00}
\end{eqnarray}
\item \underline{$Z\gamma \to HH$}
\bea
{\cal A}^0_{\pm\pm}({Z\gamma\to HH}) &=&
 \frac{-\alpha\, (s/\mws)}{16 \,\tthws}\,\,\atwozgh 
    \label{eq:zghhmm}
 \end{eqnarray}
\item \underline{$\gamma\gamma \to HH$}
\bea
{\cal A}^0_{\pm\pm}({\gamma\gamma\to HH}) & =& 
\frac{-\alpha\, (s/\mws)}{16 \,\tthws}\,\,\atwogh
     \label{eq:gghhmm}
\end{eqnarray}
\item \underline{$HH \to HH$}
\vskip 0.05cm
The amplitudes for this process do not grow with energy and hence it is not used to put any unitarity constraints.
\een
%
We have listed above the minimal set of partial wave amplitudes for a given process. partial wave amplitudes ${\cal A}^J$ for $J> 0$ are listed only when corresponding partial wave amplitudes  ${\cal A}^0$ are zero and where $J>0$ amplitudes give independent 
bound on certain couplings while $J=0$ amplitudes fail to do so. Other partial wave amplitudes that are not listed above either contain terms lower than  \ordershalf or they provide less stringent unitarity conditions involving same combination of anomalous couplings.
\end{appendix}

\bibliographystyle{spphys}       

\end{document}